\documentclass[12pt,a4paper]{article}
\usepackage{graphicx, rotating}
\usepackage{dcolumn}
\usepackage{amsmath,amssymb}
\usepackage{bm, float, cite}

\begin{document}

\title{Universality of Entanglement and Quantum Computation Complexity}
\author{Rom\'an Or\'us$^{\dag}$, Jos\'e I. Latorre$^{\dag}$
\\
\\ \emph{$^{\dag}$Dept. d'Estructura i Constituents de la Mat\`eria,}
\\ \emph{Univ. Barcelona, 08028. Barcelona, Spain.}
} 
\maketitle

\begin{abstract}

We study the universality of scaling of entanglement in Shor's
factoring algorithm and in adiabatic quantum algorithms across a
quantum phase transition for both the NP-complete Exact
Cover problem as well as the Grover's problem. 
The analytic result for Shor's algorithm shows a linear scaling of the
entropy
in terms of the number of qubits, therefore difficulting the possibility of an
efficient classical simulation protocol. 
A similar result is obtained numerically for the quantum adiabatic evolution
Exact Cover algorithm, which also shows universality
of the quantum phase transition the system evolves nearby.
 On the other hand,
entanglement in Grover's adiabatic algorithm remains a bounded
quantity even at the critical point. A classification
of scaling of entanglement appears as a natural grading of the computational
complexity of simulating quantum phase transitions.

\end{abstract}


\newpage

\section{Introduction}

One of the main theoretical challenges in quantum computation
theory is quantum algorithm design.  Some attempts to uncover underlying
principles common to all known efficient quantum algorithms have
already been explored though not definite and satisfactory answer
has been found yet. On the one hand, it has been seen that
majorization theory seems to play an important role in the
efficiency of quantum algorithms \cite{latorre, orus1, orus2}.
All known efficient quantum algorithms verify a step by step
majorization of the probability distribution associated to
the quantum register in the measurement basis. Therefore,
efficient quantum algorithms drive the system towards the final solution
by carefully reordering probability amplitudes in such a
way that a majorization arrow is always present. On
the other hand, the most relevant ingredient is likely the role
entanglement plays in quantum computational speedup. Regarding this
topic several results have recently been found \cite{ent1, ent2,
ent3, ent4, guifre, guifrebis} which suggest that entanglement is at the heart
of the power of quantum computers.

An important result has been obtained by Vidal \cite{guifre},
who proved that large entanglement of the quantum register is a
necessary condition for exponential speed-up in  quantum
computation. To be concrete, a quantum register such that the maximum Schmidt
number of any bipartition is bounded at most by a polynomial in
the size of the system can be simulated efficiently  by
classical means. The figure of merit $\chi$ proposed in \cite{guifre}
is the maximum Schmidt number of any bi-partitioning
of the quantum state or, in other words, the maximum rank of the
reduced density matrices for any possible splitting. It can be
proved that $\chi \ge 2^{E(\rho)}$, where the Von Neumann entropy
$E(\rho)$ refers to the reduced density matrix of any of the two
partitions. If $\chi = O({\rm poly}(n))$ at every step of the
computation in a quantum algorithm, then it can be efficiently
classically simulated. Exponential speed-up over
classical computation is only possible if
at some step along the computation  $\chi \sim {\rm exp}(n^a)$, or
$E(\rho) \sim n^b$, being $a$ and $b$ positive constants. In order to
exponentially accelerate the performance of classical computers
any quantum algorithm must necessarily create an exponentially
large amount of $\chi$ at some point.

Another topic of intense research concerns the behavior of
entanglement in systems undergoing a quantum phase transition
\cite{Sa99}. Quantum correlations in critical systems have been
analyzed in many situations and using a wide range of entanglement
measurements \cite{guifrebis, spin1, spin2, kike1, kike2, cirac, neill, qpt,
murg}. In particular, it has been noted \cite{kike1, kike2, neill,
qpt} that some of these measurements have important connections to
 well-known results arising from conformal field theory
\cite{cft1, cft2, cft3, korepin}. More generally,
when a splitting of a $d$-dimensional
spin system is made, the Von Neumann entropy for the reduced
density matrix of one of the subsystems $E(\rho) = -{\rm tr}(\rho
\log{ _2\rho})$ at the critical point should display a universal
leading scaling behavior determined by the \emph{area} of the
region partitioning the whole system.
This result depends on the connectivity of the Hamiltonian and
applies as is to theories with a Gaussian continuum limit.
For example, when separating
the system in the interior and the exterior of a sphere of radius
$R$ and assuming an ultraviolet cutoff $x_0$, the entropy of
$e.g.$ the interior is

\begin{equation}
E = c_1 \left( \frac{R}{x_0}\right)^{d-1} \label{cft}
\end{equation}
where $c_1$ corresponds to a known heat-kernel coefficient
\cite{cft3}. In terms of the number of spins present in the
system, this leading universal scaling behavior can be written as

\begin{equation}
E \sim n^{\frac{d-1}{d}}
\end{equation}
(which reduces to a logarithmic law for $d = 1$). This explicit
dependence of entanglement with dimensionality throws new light
into some well established results from quantum computation.

A similar situation is present in quantum adiabatic algorithms,
initially introduced by Farhi et. al. \cite{farhi1}, where the
Hamiltonian of the system depends on a control parameter $s$ which
in turn has a given time dependence. The Hamiltonians related to
adiabatic quantum computation for solving some NP-complete
problems (such as 3-SAT or Exact Cover) can be directly mapped to
interactive non-local spin systems, and therefore we can extend the
study of entanglement to include this kind of Hamiltonians. This
point of view has the additional interest of being directly
connected to the possibility of efficient classical simulations of
the quantum algorithm, by means of the protocol proposed in
ref. \cite{guifre}.

In this paper we analyze the scaling of the entropy of
entanglement in several quantum algorithms. More concretely, we
focus on Shor's quantum factoring algorithm \cite{shor}
and on a quantum algorithm by adiabatic evolution solving the
NP-complete problem Exact Cover \cite{farhi2}, finding for both of
them evidence of a quantum exponential speedup with linear scaling
of quantum correlations, which difficults the possibility of
an efficient classical simulation. We furthermore study the
adiabatic implementation of Grover's quantum search algorithm
\cite{grover, roland, vaz}, in which entanglement is a
bounded quantity even at the critical point, regardless of the size
of the system.

We have structured the paper as follows: in Sec. 2 we analytically
address the study of quantum entanglement present in Shor's
factoring algorithm. We consider the problem of universal scaling
of entanglement at the critical point of an adiabatic quantum
algorithm solving the NP-complete problem Exact Cover in Sec. 3,
where we present numerical results for systems up to 20 qubits. In
Sec. 4 we focus on the adiabatic implementation of Grover's
quantum searching algorithm, and derive analytical expressions for
the study of entanglement in the system. Finally, in Sec. 5 we
collect the conclusions of our work.

\section{Scaling of entanglement in Shor's factoring algorithm}
 It is  believed that the reason why
Shor's quantum algorithm for factorization \cite{shor}
beats so clearly its classical rivals is rooted in
the clever use it makes of quantum entanglement. Several attempts
have been made in order to understand the behavior of the quantum
correlations present along the computation \cite{ent3, ent4}. In our
case, we will concentrate in the study of the scaling
behavior for the entanglement entropy of the system. We shall first
remember both Shor's original \cite{shor} and phase-estimation
\cite{cleve} proposals of the factoring algorithm and afterwards
we shall move to the analytical analysis of their quantum correlations.

\subsection{The factoring algorithm}
The interested reader is addressed to \cite{shor, cleve, chuang,
mangel} for precise details. Given an odd integer $N$ to
factorize, we pick up a random number $a \in [1,N]$. We make the
assumption that $a$ and $N$ are Co-primes (otherwise the greatest
common divisor of $a$ and $N$ would already be a non-trivial
factor of $N$). There exists a smaller integer $r \in [1,N]$,
called the \emph{order} of the modular exponentiation $a^x \ {\rm
mod} \ N$, such that $a^r \ {\rm mod} \ N = 1$. Let us assume that
the $a$ we have chosen is such that $r$ is even and $a^{r/2} \
{\rm mod} \ N \ne -1$, which happens with very high probability
(bigger than or equal to $1/(2 \log{_2N})$). This is the case of
interest because then the greatest common divisor of $N$ and
$a^{r/2} \pm 1$ is a non-trivial factor of $N$. Therefore, the
factoring problem has been reduced to the order-finding problem of
the modular exponentiation function $a^x \ {\rm mod} \ N$, and it
is at this point where quantum mechanics comes at work. The
procedure can be casted in two different ways:

\subsubsection{Shor's proposal for order-finding}
\begin{figure}[h]
\centering
\includegraphics[angle=0, width=1\textwidth]{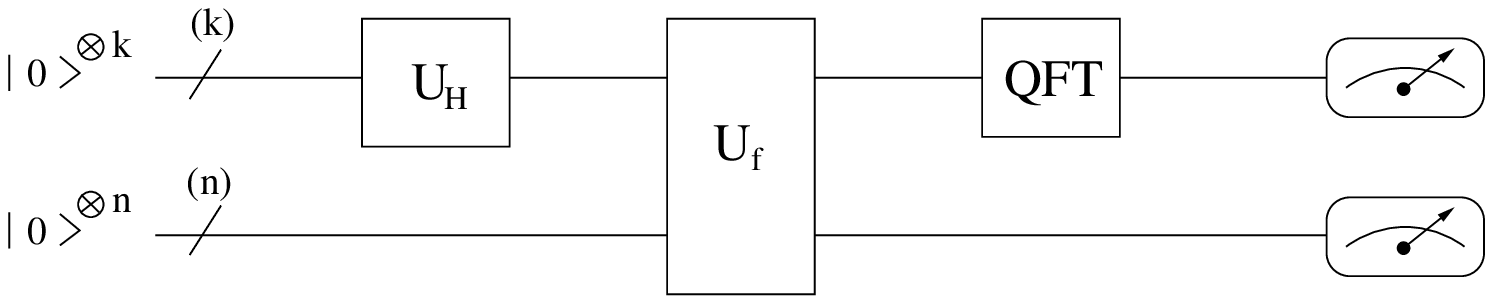}
\caption{quantum circuit for the order-finding algorithm for the
modular exponentiation function. } \label{shor}
\end{figure}

We make use of two quantum registers: a source register of $k$
qubits (such that $2^k \in [N^2, 2  N^2]$) and a target register
of $n = \lceil \log{_2N} \rceil$ qubits. The performance of the
quantum algorithm is shown in Fig. \ref{shor}, where we are making
use of the Hadamard gate initially acting over the $k$ qubits of
the source, the unitary implementation of the modular
exponentiation function
\begin{equation}
U_f |q\rangle |x\rangle = |q\rangle |(x + a^q) \ {\rm mod} \
N\rangle \label{modular}
\end{equation}
(where $|q\rangle$ and $|x\rangle$ respectively belong to the
source and target registers), and the Quantum Fourier Transform
operator
\begin{equation}
QFT |q\rangle = \frac{1}{2^{k/2}}\sum_{m=0}^{2^k-1}e^{2 \pi i q m
/ 2^k }|m\rangle \ . \label{quan}
\end{equation}
All these operations can be efficiently implemented by means of
one and two-qubit gates. Finally, a suitable classical treatment
of the final measurement of this quantum algorithm provides us
with $r$ in few steps, and therefore the prime factorization of
$N$ in a time $O((\log{_2N})^3)$.

\subsubsection{Phase-estimation proposal for order-finding}
\begin{figure}[h]
\centering
\includegraphics[angle=0, width=1\textwidth]{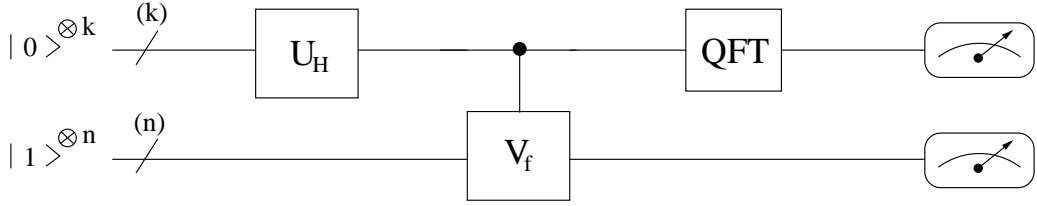}
\caption{phase-estimation version of the quantum circuit for the
order-finding algorithm. The controlled operation is $\Lambda
(V_f)$.} \label{shor2}
\end{figure}

We refer the interested reader to \cite{cleve} for more details.
The quantum circuit is similar to the one shown in the previous
section but slightly modified, as is shown in Fig. \ref{shor2}. The
unitary operator $V_f$ to which the phase-estimation procedure is
applied is defined as

\begin{equation}
V_f |x\rangle = |(a \ x) \ {\rm mod} \ N \rangle \label{modular2}
\end{equation}
(appreciate the difference between expressions (\ref{modular2})
and (\ref{modular})), being diagonalized by eigenvectors

\begin{equation}
|v_s\rangle = \frac{1}{r^{1/2}} \sum_{p = 0}^{r-1} e^{- 2 \pi i s
p / r}|a^p \ {\rm mod} \ N\rangle \label{vv}
\end{equation}
such that

\begin{equation}
V_f |v_s\rangle = e^{2 \pi i s / r}|v_s\rangle \ , \label{veigen}
\end{equation}
and satisfying the relation $\frac{1}{r^{1/2}} \sum_{s = 0}^{r-1}
|v_s\rangle = |1\rangle$. The operator is applied over the target
register being controlled on the qubits of the source in such a
way that

\begin{equation}
\Lambda (V_f) |j\rangle |x\rangle = |j\rangle V_f^j |x\rangle \ ,
\label{lambda}
\end{equation}
where by $\Lambda (V_f)$ we understand the full controlled
operation acting over the whole system, which can be efficiently
implemented in terms of one and two-qubit gates. As in the
previous case, the information provided by a final measurement of
the quantum computer enables us to get the factors of $N$ in a
time $O((\log{_2 N})^3)$.

\subsection{Analytical results}
We choose to study the amount of entanglement between the source
and the target register in the two proposed quantum circuits,
right after the modular exponentiation operation $U_f$
(Fig. \ref{shor}) or the controlled $V_f$ operation
(Fig. \ref{shor2}), and before the Quantum Fourier Transform in
both cases. At this step of the computation, the pure quantum
state of the quantum computer is easily seen to be exactly the
same for both quantum circuits, and is given by

\begin{equation}
|\psi \rangle = \frac{1}{2^{k/2}} \sum_{q = 0}^{2^k-1} |q\rangle
|a^q \ {\rm mod} \ N\rangle \ , \label{status}
\end{equation}
and therefore the density matrix of the whole system is

\begin{equation}
|\psi\rangle \langle \psi| = \frac{1}{2^k} \sum_{q, q' =
0}^{2^k-1} \left(|q\rangle \langle q'|\right) \ \left(|a^q \ {\rm
mod} \ N\rangle \langle a^{q'} \ {\rm mod} \ N|\right) \ .
\label{mat}
\end{equation}
Tracing out the quantum bits corresponding to the source, we get
the density matrix of the target register, which reads

\begin{equation}
\rho_{{\rm target}} = {\rm tr_{source}}(|\psi\rangle \langle
\psi|) = \frac{1}{2^k} \sum_{p, q, q' = 0}^{2^k-1} \left(\langle
p|q\rangle \langle q'|p\rangle\right) \ \left(|a^q \ {\rm mod} N
\rangle \langle a^{q'} \ {\rm mod} N|\right) \ ,\label{target}
\end{equation}
that is,

\begin{equation}
\rho_{{\rm target}} = \frac{1}{2^k} \sum_{p = 0}^{2^k-1} |a^p \
{\rm mod} \ N \rangle \langle a^p \ {\rm mod} \ N| \sim
\frac{1}{r} \sum_{p=0}^{r-1} |a^p \ {\rm mod} N \rangle \langle
a^p \ {\rm mod} N| \ . \label{rho}
\end{equation}
The last step comes from the fact that $a^r \ {\rm mod} \ N = 1$,
being $r \in [1, N]$ the order of the modular exponentiation. If
$2^k$ were a multiple of $r$ there would not be any approximation
and the last equation would be exact. This is not necessarily the
case, but the corrections to this expression go like $O(1/2^k)$,
thus being exponentially small in the size of the system.

It follows from expression (\ref{rho}) that the rank of the
reduced density matrix of the target register at this point of the
computation is

\begin{equation}
{\rm rank}(\rho_{{\rm target}}) = r \ . \label{ranking}
\end{equation}
Because $r \in [1, N]$, this rank is usually $O(N)$. If this were
not the case, for example if $r$ were $O(\log{_2N})$, then the
order-finding problem could be efficiently solved by a classical
naive algorithm and it would not be considered as classically
hard. Because $N$ is exponentially big in the number of qubits,
we have found a particular bipartition of the system (namely, the
bipartition between the source register and the target register)
and a step in the quantum algorithm in which the entanglement, as
measured by the rank of the reduced density matrix of one of the
subsystems, is exponentially big. This implies in turn that
Shor's quantum factoring algorithm can not be efficiently
classically simulated by any protocol in ref.\cite{guifre} owing
to the fact that at this step $\chi = O(N)$, therefore
constituting an inherent exponential quantum speed-up based on an
exponentially big amount of entanglement. It is worth noticing
that the purpose of the entanglement between the two registers
consists on leaving the source in the right periodic state to be processed
by the Quantum Fourier Transform. Measuring the register right
after the entangling gate disentangles the two registers while
leaving the source in a periodic state, and this effect can only
be accomplished by previously entangling source and target. These
conclusions apply both to Shor's original proposal (circuit of
Fig. \ref{shor}) and to the phase-estimation version (circuit of
Fig. \ref{shor2}).

The behavior of the rank of the system involves that the entropy
of entanglement of the reduced density matrix at this point will
mainly scale linearly with the number of qubits, $E \sim \log{_2 r}
\sim \log{_2 N} \sim n$, which is the hardest of all the possible scaling
laws. We will find again this strong behavior for the entropy in
Sec. 3.

\section{Scaling of entanglement in an NP-complete problem}

We now turn to analyze how entanglement scales for a quantum algorithm
based on adiabatic evolution  \cite{farhi1}, designed to solve
 the NP-complete problem Exact Cover \cite{farhi2}.  We first briefly
 review the proposal and, then,  we
consider the study of the properties of the system, in particular
the behavior of the entanglement entropy for a given bipartition
of the ground state.

\subsection{Adiabatic quantum computation}

The adiabatic model of quantum computation deals with the problem
of finding the ground state of a given system represented by its
Hamiltonian. Many relevant computational problems (such as 3-SAT)
can be mapped to this situation. The method is briefly summarized
as follows: we start from a time dependent Hamiltonian of the form

\begin{equation}
H(s(t)) = (1-s(t)) H_0 + s(t) H_p \ , \label{ham}
\end{equation}
where $H_0$ and $H_p$ are the initial and problem Hamiltonian
respectively, and $s(t)$ is a time-dependent function satisfying
the boundary conditions $s(0) = 0$ and $s(T) = 1$ for a given $T$.
The desired solution to a certain problem is codified in the
ground state of $H_p$. The gap between the ground and the first
excited state of the instantaneous Hamiltonian at time $t$ will be
called $g(t)$. Let us define $g_{min}$ as the global minimum of
$g(t)$ for $t$ in the interval $[0, T]$. If at time $T$ the ground
state is given by the state $|E_0; T\rangle$, the adiabatic
theorem states that if we prepare the system in its ground state
at $t=0$ (which is assumed to be easy to prepare) and let it
evolve under this Hamiltonian, then

\begin{equation}
|\langle E_0; T|\psi(T)\rangle|^2 \geq 1 - \epsilon^2
\label{probab}
\end{equation}
provided that

\begin{equation}
\frac{{\rm max} |\frac{dH_{1,0}}{dt}|}{g^2_{min}} \leq \epsilon
\label{cond}
\end{equation}
where $H_{1,0}$ is the Hamiltonian matrix element between the
ground and first excited state, $\epsilon << 1$, and the
maximization is taken over the whole time interval $[0,T]$.
Because the problem Hamiltonian codifies the solution to the
problem in its ground state, we get the desired solution with high
probability after a time $T$. A closer look to the adiabatic
theorem tells us that $T$ dramatically depends on the scaling of
the inverse of $g_{min}^2$ with the size of the system. More
concretely, if the gap is only polynomially small in the number of
qubits (that is to say, it scales as $O (1/{\rm poly}(n))$, the
computational time is $O({\rm poly}(n))$, whereas if the gap is
exponentially small ($O(2^{-n})$) the algorithm makes use of an
exponentially big time to reach the solution.

The explicit functional dependence of the parameter $s(t)$ on time
can be very diverse. The point of view we adopt in the present
paper is such that this time dependence is not taken into account,
as we study the properties of the system as a function of $s$,
which will be understood as the Hamiltonian parameter. We will in
particular analyze the entanglement properties of the ground state
of $H(s)$, as adiabatic quantum computation assumes that the
quantum state remains always close to the instantaneous ground
state of the Hamiltonian all along the computation. Note that we
are dealing with a system which is suitable to undergo a quantum
phase transition at some critical value of the Hamiltonian
parameter, and therefore we expect to achieve the biggest quantum
correlations at this point. The question is how this big quantum
correlations scale with the size of the system when dealing with
interesting problems. This is the starting point for the next two
sections.

\subsection{Exact Cover}

The NP-complete problem Exact Cover is a particular case of the
3-SAT problem, and is defined as follows: given the $n$ boolean
variables $\{x_i\}_{i=1,\ldots n}$, $x_i = 0,1 \ \forall \ i$,
where $i$ is regarded as the bit index, we define a \emph{clause}
of Exact Cover involving the three qubits $i$, $j$ and $k$  (say,
clause ``$C$") by the equation $x_i + x_j + x_k = 1$. There are
only three assignments of the set of variables $\{x_i, x_j, x_k
\}$ that satisfy this equation, namely, $\{1,0,0\}$, $\{0,1,0\}$
and $\{0,0,1\}$. The clause can be more specifically expressed in
terms of a boolean function in Conjunctive Normal Form (CNF) as
\begin{eqnarray}
\phi_{C}(x_i,x_j,x_k) &&= (x_i \lor x_j \lor x_k)\land(\neg x_i
\lor \neg x_j \lor \neg x_k)\land(\neg x_i \lor \neg x_j \lor
x_k) \nonumber \\
&&\land(\neg x_i \lor x_j \lor \neg x_k)\land(x_i \lor \neg x_j
\lor \neg x_k) \ , \label{CNF}
\end{eqnarray}
so $\phi_{C}(x_i,x_j,x_k) = 1$ as long as the clause is properly
satisfied. An \emph{instance} of Exact Cover is a collection of
clauses which involves different groups of three qubits. The
problem is to find a string of bits $\{x_1, x_2 \ldots , x_n \}$
which satisfies all the clauses.

This problem can be mapped into finding the ground state of a
Hamiltonian $H_p$ in the following way: given a clause $C$ define
the Hamiltonian associated to this clause as

\begin{eqnarray}
H_C &=&
\frac{1}{2}(1+\sigma_i^z)\frac{1}{2}(1+\sigma_j^z)\frac{1}{2}(1+\sigma_k^z)
 \nonumber \\ &+&
\frac{1}{2}(1-\sigma_i^z)\frac{1}{2}(1-\sigma_j^z)\frac{1}{2}(1-\sigma_k^z)
 \nonumber \\ &+&
\frac{1}{2}(1-\sigma_i^z)\frac{1}{2}(1-\sigma_j^z)\frac{1}{2}(1+\sigma_k^z)
 \nonumber \\ &+&
\frac{1}{2}(1-\sigma_i^z)\frac{1}{2}(1+\sigma_j^z)\frac{1}{2}(1-\sigma_k^z)
 \nonumber \\ &+&
\frac{1}{2}(1+\sigma_i^z)\frac{1}{2}(1-\sigma_j^z)\frac{1}{2}(1-\sigma_k^z)
 \ , \label{ham}
\end{eqnarray}
where we have defined $\sigma^z |0\rangle = |0\rangle$, $\sigma^z
|1\rangle = -|1\rangle$. Note the parallelism between equations
(\ref{CNF}) and (\ref{ham}). The quantum states of the
computational basis that are eigenstates of $H_C$ with zero
eigenvalue (ground states) are the ones that correspond to the bit
string which satisfies $C$, whereas the rest of the computational
states are penalized with an energy equal to one. Now, we
construct the problem Hamiltonian as the sum of all the
Hamiltonians corresponding to all the clauses in our particular
instance, that is to say,

\begin{equation}
H_p = \sum_{C \ \in \ {\rm instance}} H_C \ , \label{hamil}
\end{equation}
so the ground state of this Hamiltonian corresponds to the quantum
state whose bit string satisfies \emph{all} the clauses. We have
reduced the original problem stated in terms of boolean logic to
the hard task of finding the ground state of a two and three body
interactive spin Hamiltonian with local magnetic fields. Observe that
the couplings depend on the particular instance we are dealing
with, and that the spin system has not an a priori well defined
dimensionality neither a well defined lattice topology, in
contrast with some usual simple spin models.

We now define our s-dependent Hamiltonian $H(s)$ as a linear
interpolation between an initial Hamiltonian $H_0$ and $H_p$:

\begin{equation}
H(s) = (1-s)H_0 + s H_p \label{finalham}
\end{equation}
where we take the initial Hamiltonian $H_0$ to be basically a
magnetic field in the $x$ direction, more concretely,

\begin{equation}
H_0 = \sum_{i = 1}^n \frac{d_i}{2}(1-\sigma_i^x) \ , \label{h0}
\end{equation}
where $d_i$ is the number of clauses in which qubit $i$ appears,
and $\sigma^x |+\rangle = |+\rangle$, with $|+\rangle =
\frac{1}{\sqrt{2}}(|0\rangle + |1\rangle)$, so the ground state of
$H_0$ is an equal superposition of all the possible computational
states. Observe that $H(s)$ is, apart from a constant factor, a
sum of terms involving local magnetic fields in the $x$ and $z$
direction, together with two and three-body interaction coupling
terms in the $z$ component. This system is suitable to undergo a
quantum phase transition (in the limit of infinite $n$) as $s$ is
shifted from $0$ to $1$. The study of this phenomena is the aim of
the following section.

\subsection{Numerical results up to 20 qubits}

We have randomly generated instances for Exact Cover with only one
possible satisfying assignment and have constructed the
corresponding problem Hamiltonians. Instances are produced by
adding clauses at random until there is exactly one satisfying
assignment, starting over if we end up with no satisfying
assignments. According to \cite{farhi2}, these are believed to be
the most difficult instances for the adiabatic algorithm. Our
analysis proceeds as follows:

\bigskip

{\bf \emph{Appearance of a quantum phase transition}}

\begin{figure}[h]
\centering
\includegraphics[angle=-90, width=0.8\textwidth]{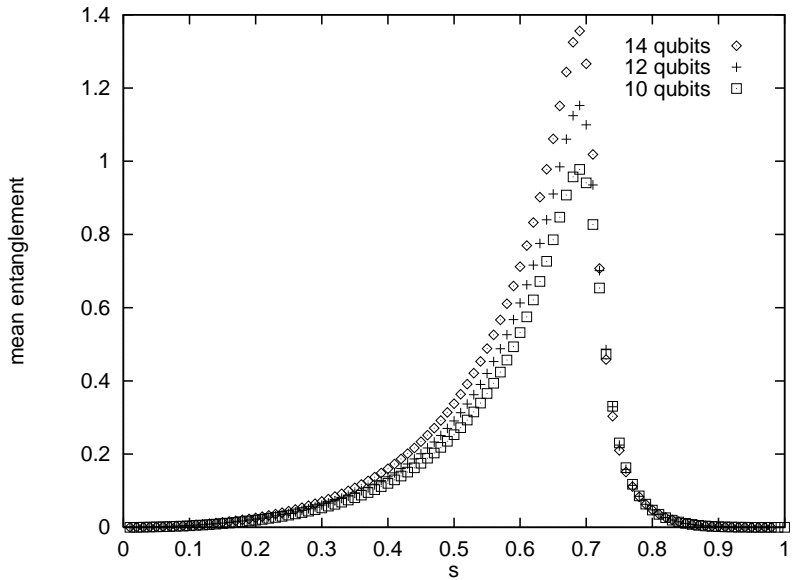}
\caption{evolution of the entanglement entropy between the two
blocks of size $n/2$ when a bipartition of the system is made, on
average over 300 different instances with one satisfying
assignment. A peak in the correlations appears for $s_c \sim 0.7$
in the three cases.} \label{mix-300}
\end{figure}

We have generated 300 Exact Cover instances (300 random
Hamiltonians with a non-degenerated ground state) and have
calculated the ground state for 10, 12 and 14 qubits for different
values of the parameter $s$ in steps of $0.01$. We then consider a
particular bipartition of the system into two blocks of $n/2$
qubits, namely, the first $n/2$ qubits versus the rest, and have
calculated the entanglement entropy between the two blocks. For
each of the randomly generated Hamiltonians we observe a peak in
the entanglement entropy around a critical value of the parameter
$s_c \sim 0.7$. We have averaged the obtained curves over the 300
instances and have obtained the plot from Fig. \ref{mix-300}.

\begin{figure}[h]
\centering
\includegraphics[angle=-90, width=0.8\textwidth]{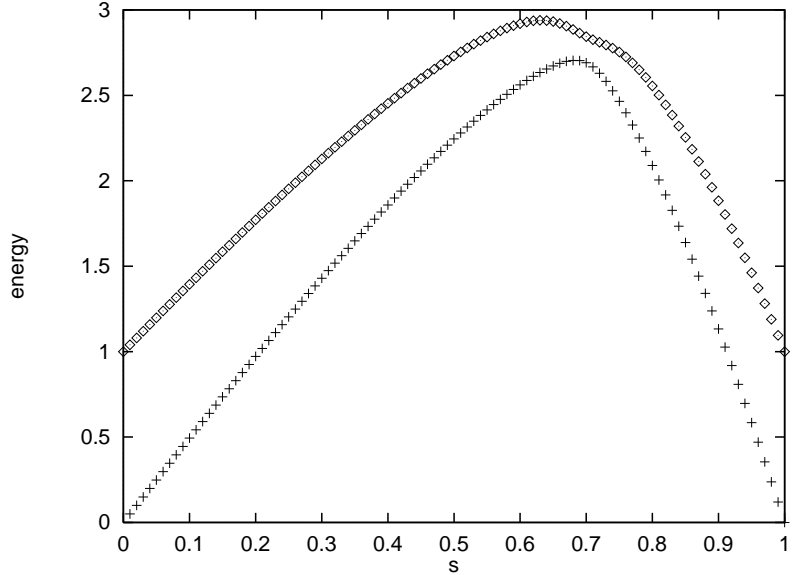}
\caption{energies of the ground state and first excited state for
a typical instance with one satisfying assignment of Exact Cover
in the case of 10 qubits (in dimensionless units). The energy gap approaches its minimum at $s_c
\sim 0.7$.} \label{gap-10}
\end{figure}

The point at which the entropy of entanglement reaches its maximum
value is identified as the one corresponding to the critical point
of a quantum phase transition in the system (in the limit of
infinite size). This interpretation is reinforced by the
observation of the typical energy eigenvalues of the system. For a
typical instance of 10 qubits we observe that the energy gap
between the ground state and the first excited state reaches a
minimum precisely for a value of the parameter $s_c \sim 0.7$ (see
Fig. \ref{gap-10}).

We observe from Fig. \ref{mix-300} that the peak in the entropy is
highly asymmetric with respect to the parameter $s$. A detailed
study of the way this peak seems to diverge near the critical
region seems to indicate that the growth of entanglement is slower
at the beginning of the evolution and fits remarkably well a curve
of the type $E \sim \log{| \log{(s-s_c)}|}$, whereas the falling
down of the peak is better parameterized by a power law $E \sim
|s-s_c|^{-\alpha}$ with $\alpha \sim 2.3$, being $\alpha$ a
certain critical exponent. These laws governing the critical
region fit better and better the data as the number of qubits is
increased.

\bigskip

{\bf \emph{Analysis of different bi-partitions of the system}}

\begin{figure}
\centering
\includegraphics[angle=-90, width=0.8\textwidth]{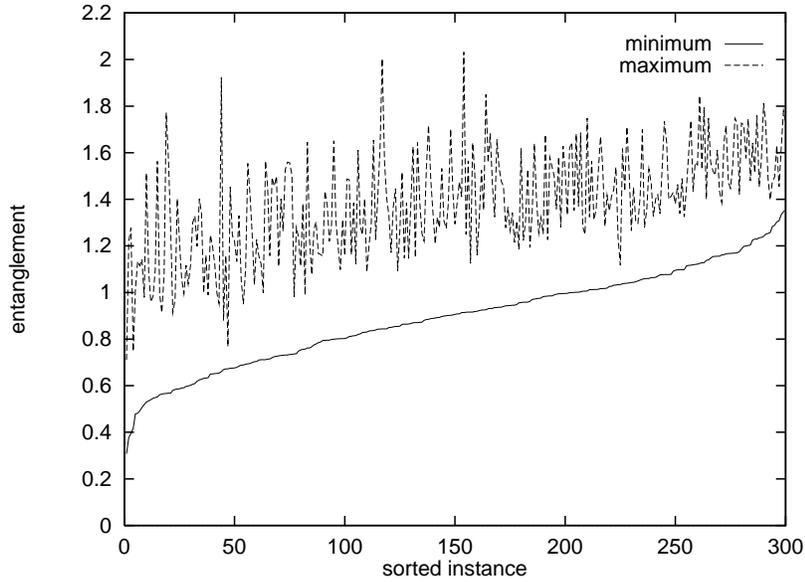}
\caption{minimum and maximum entropy over all possible
bi-partitions of a $10$-qubit system for each of the $300$
randomly generated instances of Exact Cover. Instances are sorted
such that the minimum entanglement monotonically increases.}
\label{part1}
\end{figure}

Explicit numerical analysis for $10$ qubits tells us that all
possible bi-partitions for each one of the instances produce
entropies at the critical point of the same order of magnitude -as
expected from the non-locality of the interactions-. This is
represented in Fig. \ref{part1}, where we plot the minimum and
maximum entanglement obtained from all the possible partitions of
the system for each one of the generated instances (points are
sorted such that the minimum entropy monotonically increases).

\begin{figure}[H]
\centering
\includegraphics[angle=-90, width=0.8\textwidth]{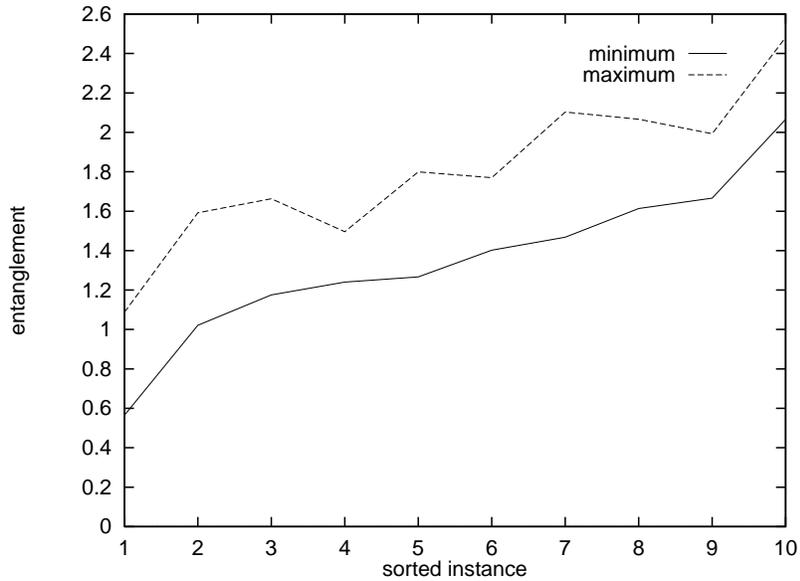}
\caption{minimum and maximum entropy over $64$ bi-partitions of a
$16$-qubit system for $10$ randomly generated instances of Exact
Cover. Instances are sorted such that the minimum entanglement
monotonically increases.} \label{part2}
\end{figure}

Similar conclusions derive from the data plotted in
Fig. \ref{part2}, where we have considered again the same
quantities but looking at $64$ partitions of the ground state for
$10$ different instances of $16$ qubits. According to these
results we restrict ourselves in what follows to the analysis of a
particular bipartition of the system, namely the first $n/2$
qubits versus the rest.

It is worth emphasizing that the existence of a single partition
with exponentially large entanglement makes the algorithm not amenable
to classical simulation. The above result is stronger and shows that
essentially all partitions are highly entangled. The system
is definitely hard to simulate by classical means.

\bigskip

{\bf \emph{Scaling laws for the minimum energy gap and the entanglement
entropy}}

\begin{figure}
\centering
\includegraphics[angle=-90, width=0.8\textwidth]{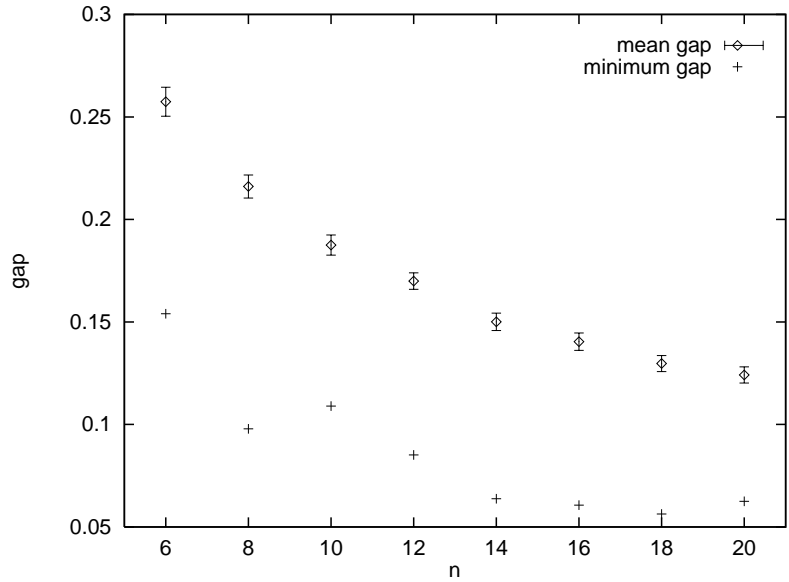}
\caption{scaling of the minimum energy gap (in dimensionless units) with the size of the system,
both in the worst case and in the mean case over all the randomly
generated instances. Error bars give $95$ per cent of confidence
level for the mean.} \label{gap}
\end{figure}

\begin{figure}
\centering
\includegraphics[angle=-90, width=0.8\textwidth]{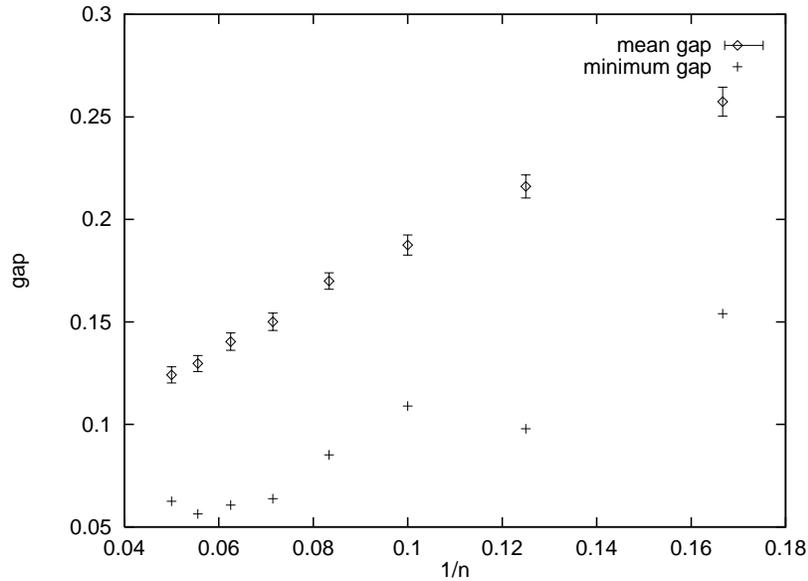}
\caption{minimum energy gap (in dimensionless units) versus the inverse size of the system, both
in the worst case and in the mean case over all the randomly
generated instances. Error bars give $95$ per cent of confidence
level for the mean. The behavior is apparently linear. }
\label{gap2}
\end{figure}

To characterize the finite-size
behavior of the quantum phase transition, we have generated 300
random instances of Exact Cover with only one satisfying
assignment from 6 to 20 qubits, and studied the maximum Von
Neumann entropy for a bipartition of the system as well as the
minimum gap, both in the worst case and in the mean case over all
the randomly generated instances. We must point out that the scaling
laws found in this section are limited to the small systems we can handle with
in our computers. Increasing the number of qubits may lead to corrections in
the numerical results, which should be of particular importance for a more precise
time-complexity analysis of the adiabatic algorithm.   
 Fig. \ref{gap} represents the
behavior of the gap in the worst and mean cases. From
Fig. \ref{gap2} it is noticed that the gap seems to obey a scaling
law of the style $O(1/n)$, being $n$ the number of qubits, which
would assure a polynomial-time quantum computation. This law is in
agreement with the results in \cite{farhi2}, and are in concordance with 
the idea that the energy gap typically vanishes as the inverse of the volume
in condensed matter systems (here the volume is the number of qubits). 
Error bars in the two plots give $95$ per cent of confidence level in the numerically
calculated mean.

\begin{figure}
\centering
\includegraphics[angle=-90, width=0.8\textwidth]{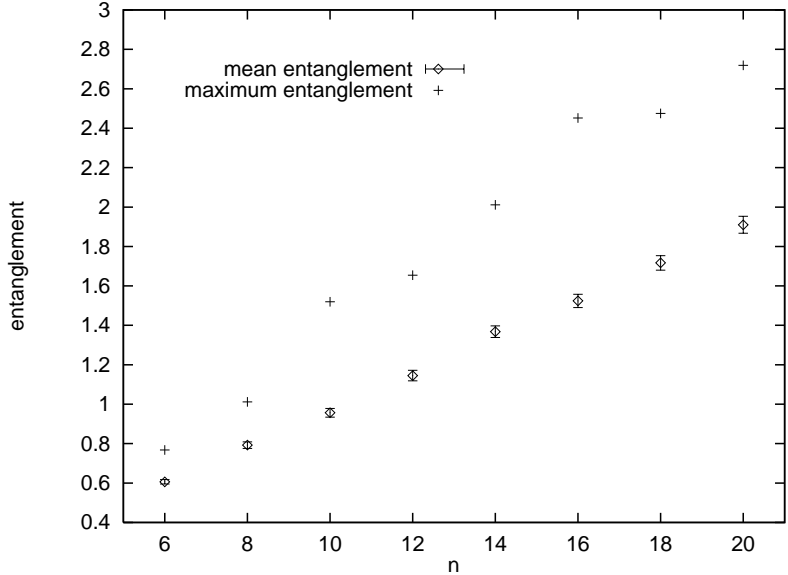}
\caption{scaling of the entanglement entropy for an equally sized
bipartition of the system, both in the worst case and in the mean
case over all the randomly generated instances. Error bars give
$95$ per cent of confidence level for the mean. The data are
consistent with a linear scaling. } \label{ent}
\end{figure}

We have considered as well the scaling behavior of the
entanglement entropy for an equally sized bipartition of the
system also in the worst and in the mean case. The obtained data
from our simulations are plotted in Fig. \ref{ent} -where error
bars give $95$ per cent of confidence level in the mean- and seem
to be in agreement with a strongly linear scaling of entanglement
as a function of the size of the number of qubits. More
concretely, a numerical linear fit for the mean entanglement
entropy gives us the law $E \sim .1 \ n$. Observe that the entropy
of entanglement does not get saturated in its maximum allowed
value (which would be $E = n/2$ for $n$ qubits), so we can say
that only a twenty percent of all the possible potential available
entanglement appears in the quantum algorithm. Linearity in the
scaling law would imply that this quantum computation by adiabatic
evolution, after a suitable discretization of the continuous time
dependence, could not be classically simulated by the protocol of
ref. \cite{guifre}. Given that the scaling of the gap seems to
indicate that the quantum computation runs in a polynomial time in
the size of the system, our conclusion is that apparently we are
in front of an exponentially fast quantum computation that seems
extremely difficult (if not impossible) to be efficiently
simulated by classical means. This could be an inherent quantum
mechanical exponential speedup that can be understood in terms of
the linear scaling of the entropy of entanglement. Note also the
parallelism with the behavior of the entanglement found in Shor's
algorithm in Sec. 2. As a remark, our numerical analysis shows that the quantum
algorithm is difficult to be simulated classically in an efficient way, which does not
necessarily imply that the quantum computer runs exponentially faster than the
classical one, as our time-complexity analysis is limited to 20 qubits. 


The linear behavior for the entropy with respect to the size of
the system could in principle be expected according to the
following qualitative reasoning. Naively, the entropy was expected
to scale as the area of the boundary of the splitting, according
to some considerations taken from conformal field theory (see
\cite{kike1, kike2, cft1, cft2, cft3}). This area-law is in some 
sense natural: because the entropy 
value is the same for both density matrices arising from the two subsystems, it
can only be a function of their shared properties, and these are geometrically
encoded in the area of the common boundary. For a system of $n$
qubits, this implies a scaling law for the entropy like $E \sim
n^{(d-1)/d}$ (which reduces to a logarithm for $d = 1$). Our
system does not have a well defined dimensionality, but owing to
the fact that there are many random two and three body
interactions, the effective (fractal) dimensionality of the system
should be very large. Therefore, we expect a linear (or almost
linear) scaling, which is what we have numerically obtained. The
data seems to indicate that such an effective dimensionality is
around $d \sim n$, thus diverging as $n$ goes to infinity.

It is possible to compare our apparently linear scaling of
the mean entropy of entanglement with the known results obtained by averaging
this quantity over the entire manifold of $n$-qubit pure states, with respect
to the natural Fubini-Study measure. According to the results conjectured by Page
\cite{page} and later proved in \cite{sen}, the average entropy for an
equally-sized bipartition of a random $n$-qubit pure state in the large $n$
limit can be approximated by $E \sim (n/2) - 1/(2 \ \ln{2})$ (in our
notation), therefore displaying as well a linear scaling law (but different
from ours). In fact, this is an indicator that most of the $n$-qubit pure
states are highly entangled, and that adiabatic quantum computation naturally
brings the system close to these highly entangled regions of the pure state
manifold (more information about the average entanglement of an $n$-qubit
system can be found in \cite{viv}).

\begin{figure}
\centering
\includegraphics[angle=-90, width=0.8\textwidth]{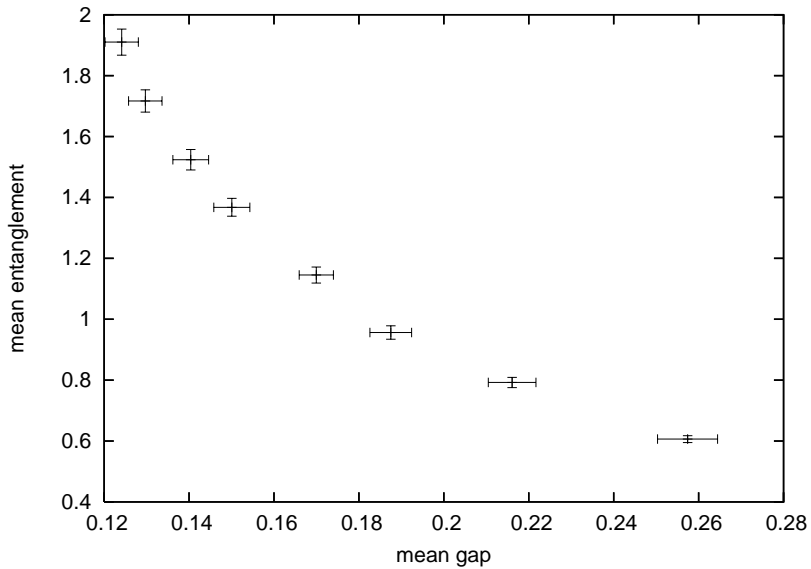}
\caption{mean entropy of entanglement versus mean size of the energy gap (in
  dimensionless units).
Error bars give $95$ per cent of confidence level for the means.
Each point corresponds to a fixed number of qubits. }
\label{entgap1}
\end{figure}

\begin{figure}
\centering
\includegraphics[angle=-90, width=0.8\textwidth]{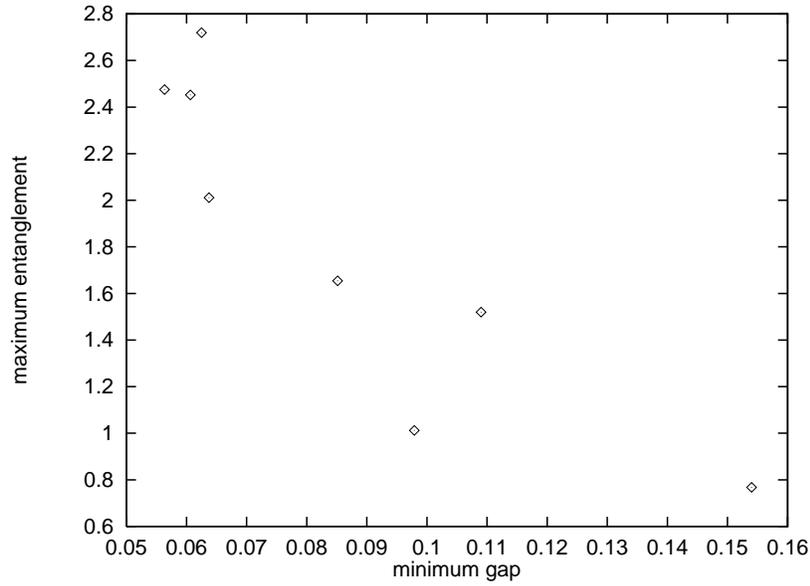}
\caption{maximum entropy of entanglement versus minimum size of
the energy gap (in dimensionless units). Each point corresponds to a fixed number of qubits.}
\label{entgap2}
\end{figure}

\bigskip

{\bf \emph{The entanglement-gap plane}}

The plots in
Fig. \ref{entgap1} and Fig. \ref{entgap2} show the behavior of the peak
in the entanglement versus the gap, both again in the average and
the worst case for all the generated instances. Clearly, as the
gap becomes smaller the production of entanglement in the
algorithm increases. A compression of the energy levels correlates with
high quantum correlations in the system.

\bigskip

{\bf \emph{Convergence of the critical points}}

\begin{figure}
\centering
\includegraphics[angle=-90, width=0.8\textwidth]{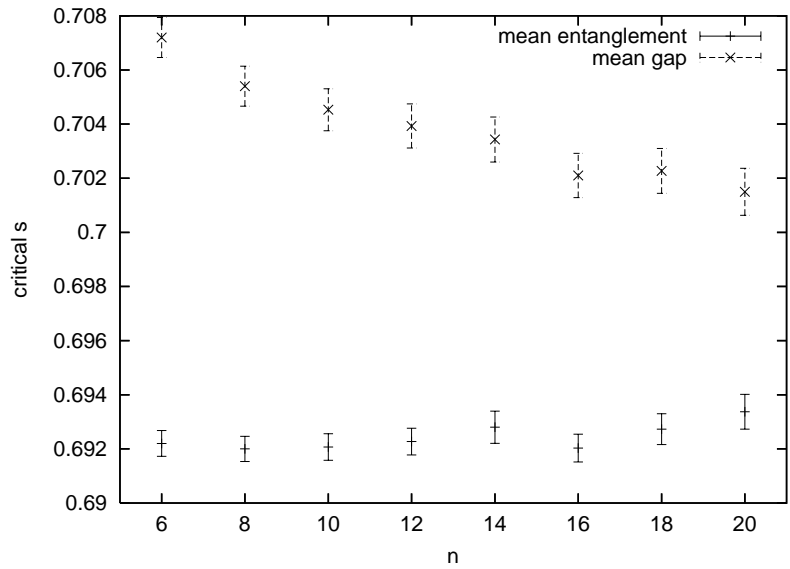}
\caption{mean critical point for the energy gap and for the entropy.
Error bars give $95$ per cent of confidence level for the means.
Note that they tend to approach as the size of the system is
increased.} \label{move}
\end{figure}

The critical point $s_c$ seems to be bounded by the values of $s$
associated to the minimum gap
and to the maximum entropy.
Actually, the  critical point
corresponding to the minimum size of the energy gap is
systematically slightly bigger than the critical point
corresponding to the peak in the entropy. By increasing the size
of the system these two points  converge towards the same
value, which would correspond to the true critical point of a
system of infinite size. This effect is neatly observed in
Fig. \ref{move}, which displays the values of $s$ associated
to  the mean critical
points both for the gap and for the entropy as a function of $n$.

\bigskip

{\bf \emph{Universality}}

All the above results suggest that the system comes
close to  a quantum phase transition.
The characterization we have presented based on the study of
averages over instances reconstructs its universal behavior. Results do not
depend on particular microscopic details of the Hamiltonian, such
as the interactions shared by the spins or the strength of local
magnetic fields. Any adiabatic algorithm solving a $k$-sat problem
and built in the same way we have done for Exact Cover should
display on average exactly the same properties we have found
\emph{regardless of the value of $k$}, which follows from
universality (the case $k=2$, though not being NP-complete, should display
also this property as its hamiltonian would consist as well of local
interactions in a big-dimensional lattice; $k=1$ is a particular case, as its
hamiltonian is non-interacting). Linear scaling of entanglement should therefore be a
universal law for these kind of quantum algorithms. The
specific coefficients of the scaling law for the entropy should be a function
only of the connectivity of the system, that is on the type
of clauses defining the instances.

\begin{figure}[h]
\centering
\includegraphics[angle=-90, width=0.8\textwidth]{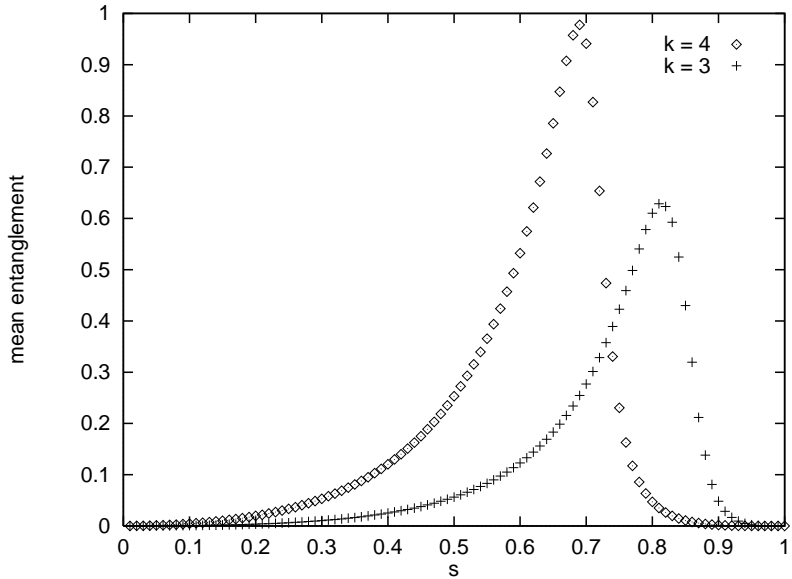}
\caption{entanglement as a function of the Hamiltonian parameter
for clauses of Exact Cover involving $3$ ($k = 3$) and $4$ ($k=4$)
qubits, for a $10$-qubit system, averaged over all the randomly generated
instances.} 
\label{univ1}
\end{figure}

We have explicitly checked this assertion by numerical simulations
for clauses of Exact Cover but involving $4$ qubits ($x_i + x_j +
x_k + x_l = 1$), which is a particular case of $4$-sat. In
Fig. \ref{univ1} we plot the behavior of the entropy of
entanglement for a $10$-qubit system for these type of clauses and
compare it to the same quantity calculated previously for the
clauses involving $3$ qubits (the usual Exact Cover Hamiltonian).
We observe again the appearance of a peak in the entropy, which
means that the system is evolving close to a quantum phase
transition.

\begin{figure}
\centering
\includegraphics[angle=-90, width=0.8\textwidth]{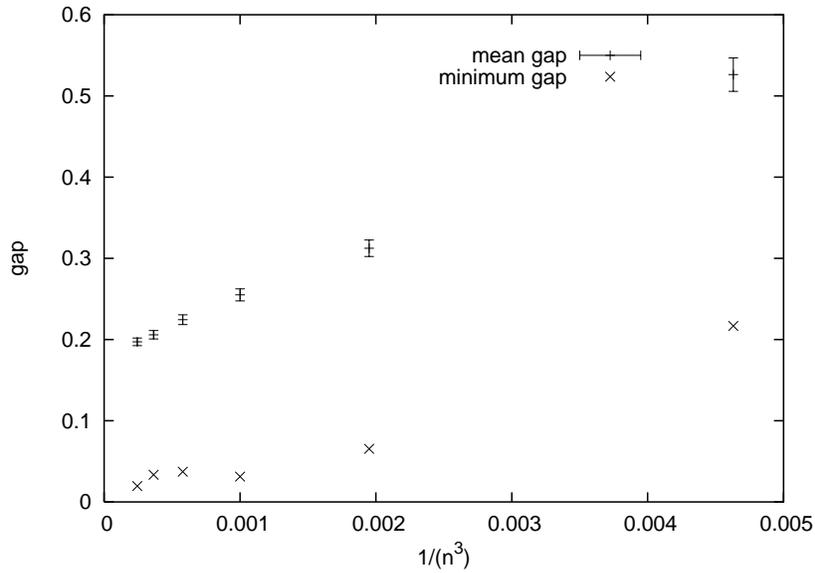}
\caption{minimum energy gap (in dimensionless units) versus
 $1/(n^3)$,
both in the worst and in the mean cases over all the randomly
generated instances of clauses involving $4$ qubits, up to
$n = 16$. Error bars give $95$ per cent of confidence
level for the mean. The behavior seems to be linear.}
\label{univ3}
\end{figure}
\begin{figure}[H]
\centering
\includegraphics[angle=-90, width=0.8\textwidth]{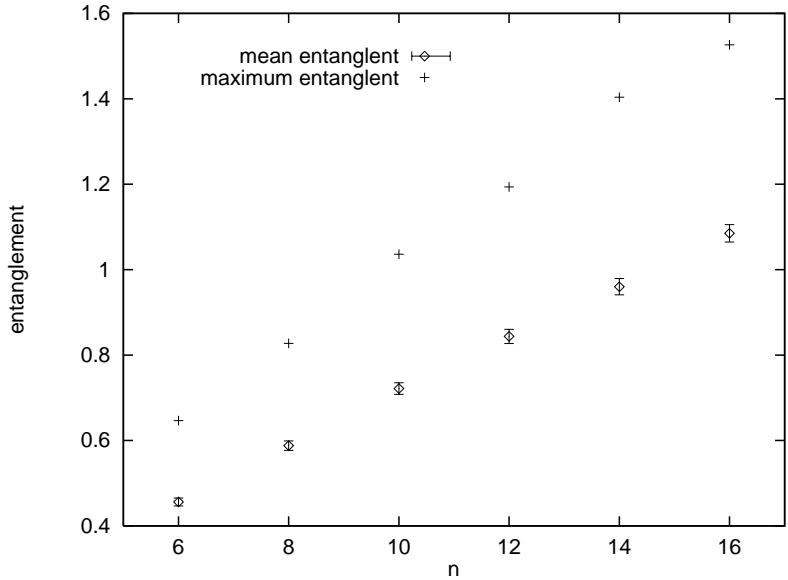}
\caption{scaling of the entanglement entropy for an equally sized
bipartition of the system, both in the worst and in the mean
cases over all the randomly generated instances of clauses
involving $4$ qubits, up to $n = 16$. Error bars give
$95$ per cent of confidence level for the mean. The data are
consistent with a linear scaling. } \label{univ2}
\end{figure}

Figures \ref{univ3} and \ref{univ2} respectively show the scaling
of the energy gap in the mean and worst case and the scaling of
the peak in the entropy in the mean and worst case as well, up to
$16$ qubits. Error bars give again $95$ per cent of confidence
level for the means. The behavior is similar to the one already
found for the instances of Exact Cover involving $3$ qubits
(figures \ref{gap2} and \ref{ent}), which supports the idea of the
universality of the results. The minimum energy gap seems to scale
in this case as $\sim \frac{1}{n^3}$ ($n$ being the number of
qubits), which would guarantee again a polynomial-time quantum
adiabatic evolution.

\section{Scaling of entanglement in adiabatic Grover's algorithm}

Let us now consider the adiabatic implementation of Grover's quantum
searching algorithm in terms of a Hamiltonian evolution
\cite{grover, roland, vaz} and study its properties as a function
of the number of qubits and the parameter $s$. For this problem, it is
possible to
compute all the results analytically, so we shall get a closed
expression for the scaling of entanglement. As a side remark,
 it is worth noting
that the treatment made in
\cite{guifre} is not valid for oracular problems as it is assumed
that all quantum gates are known in advanced. Independently of
this issue, we shall see that the system remains little entangled
between calls to the oracle.

\subsection{Implementation of Grover's searching algorithm with
adiabatic quantum computation}

Grover's searching algorithm \cite{grover} can be implemented in
adiabatic quantum computation by means of the $s$-dependent
Hamiltonian

\begin{equation}
H(s) = (1-s)(I - |s\rangle \langle s|) + s (I-|x_0\rangle \langle
x_0|) \ , \label{grahm}
\end{equation}
where $|s\rangle \equiv \frac{1}{2^{n/2}} \sum_{x = 0}^{2^n-1}
|x\rangle$, $n$ is the number of qubits, and $|x_0\rangle$ is the
marked state. The computation takes the quantum state from an
equal superposition of all  computational states
directly to the state $|x_0\rangle$, as long as the evolution
remains adiabatic. The time the algorithm takes to succeed depends
on how we choose the parameterization of $s$ in terms
of time. Our aim is to compute the amount of entanglement 
present in the register and need not deal with the explicit
dependence of the parameter $s$ on 
time and its consequences (see \cite{roland, vaz}
for further information about this topic).

It is straightforward to check that the Hamiltonian (\ref{grahm})
has its minimum gap between the ground and first excited states at
$s = 0.5$, which goes to zero exponentially fast as the number of
qubits in the system is increased. Therefore, this Hamiltonian
apparently seems to undergo a quantum phase transition in the
limit of infinite size at $s=0.5$. 
Quantum correlations approach their  maximum
for this value of $s$ (for more on Grover's problem as a quantum
phase transition, see \cite{childs}).

\subsection{Analytical results}

It can be seen (see for example \cite{expo}) that the ground state
energy of the Hamiltonian given in equation (\ref{grahm})
corresponds to the expression

\begin{equation}
E_-(s) = \frac{1}{2} \left(1-\sqrt{(1-2s)^2
+\frac{4}{2^n}s(1-s)}\right) \ , \label{ground}
\end{equation}
being $s$ is the Hamiltonian parameter. The corresponding
normalized ground state eigenvector is given by

\begin{equation}
|E_-(s)\rangle = a |x_0\rangle + b \sum_{x \neq x_0} |x\rangle \ ,
\label{groundstate}
\end{equation}
where we have defined the quantities

\begin{eqnarray}
a &\equiv& \alpha \ b \nonumber \\
&& \nonumber \\
b^2 &\equiv& \frac{1}{2^n - 1 + \alpha^2} \nonumber \\
&& \nonumber \\
\alpha &\equiv&
\frac{2^n-1}{2^n-1-\left(\frac{2^n}{1-s}\right)E_-(s)} \ .
\label{expressions}
\end{eqnarray}

In all the forthcoming analysis we will assume that the marked
state corresponds to $|x_0\rangle = |0\rangle$, which will not
alter our results. The corresponding density matrix for the ground
state of the whole system of $n$ qubits is then given by

\begin{equation}
\rho_n = b^2(\alpha^2-2\alpha +1)|0\rangle\langle 0| + b^2
|\phi\rangle \langle \phi| + b^2(\alpha-1)(|\phi\rangle\langle 0|
+ |0\rangle \langle \phi|) \ , \label{density}
\end{equation}
where we have defined $|\phi\rangle$ as the the unnormalized sum
of all the computational quantum states (including the marked
one), $|\phi \rangle \equiv \sum_{x = 0}^{2^n-1} |x\rangle$.
Taking the partial trace over half of the qubits, regardless of
what $n/2$ qubits we choose, we find the reduced density matrix

\begin{equation}
\rho_{n/2} = b^2(\alpha^2 - 2 \alpha + 1)|0' \rangle \langle 0'| +
2^{n/2} b^2 |\phi'\rangle \langle \phi'| + b^2 (\alpha - 1)
(|\phi'\rangle \langle 0'| + |0'\rangle \langle \phi'|) \ ,
\label{densityn2}
\end{equation}
where we understand that $|0'\rangle$ is the remaining marked
state for the subsystem of $n/2$ qubits and $|\phi'\rangle \equiv
\sum_{x=0}^{2^{n/2}-1} |x\rangle$ is the remaining unnormalized
equally superposition of all the possible computational states for
the subsystem. Defining the quantities

\begin{eqnarray}
A &\equiv& \frac{\alpha^2 + 2^{n/2} - 1}{\alpha^2 + 2^n - 1}
\nonumber \\
&& \nonumber \\
B &\equiv& \frac{\alpha + 2^{n/2} -1}{\alpha^2 + 2^n - 1}
\nonumber \\
&& \nonumber \\
C &\equiv& \frac{2^{n/2}}{\alpha^2 + 2^n - 1} \label{definitions}
\end{eqnarray}
(note that $A + (2^{n/2} - 1) C = 1$), the density operator for
the reduced system of $n/2$ qubits can be expressed in matrix
notation as

\begin{equation}
\rho_{n/2} =
\begin{pmatrix}
  A & B & \cdots & B \\
  B & C & \cdots & C \\
  \vdots & \vdots & \ddots & \vdots \\
  B & C & \cdots & C
\end{pmatrix} \ ,
\label{matrix}
\end{equation}
where its dimensions are $2^{n/2} \times 2^{n/2}$. We clearly see
that the density matrix has rank equal to $2$. Therefore, because
${\rm rank}(\rho) \ge 2^{E(\rho)} \ \forall \rho$ (where $E(\rho)$
is the Von Neumann entropy of the density matrix $\rho$) we
conclude that $E(\rho_{n/2})$, which corresponds to our
entanglement measure between the two blocks of qubits, is always
$\le 1$. This holds true even for non symmetric bi-partitions of
the complete system. Regardless of the number of qubits,
entanglement in Grover's adiabatic algorithm is always a
\emph{bounded} quantity for any $s$, in contrast with the results
obtained in the previous sections for Shor's factoring algorithm
and for the Exact Cover problem. Grover's adiabatic quantum
algorithm essentially makes use of very little entanglement, but even
this bounded quantity of quantum correlations is enough to give a
square root speedup.

We have explicitly calculated the Von Neumann entropy for
$\rho_{n/2}$. Because the rank of the reduced density matrix is
two, there are only two non-vanishing eigenvalues that contribute
in the calculation which are

\begin{equation}
\lambda_{\pm} = \frac{1}{2}\left(1 \pm \sqrt{1 - 4 (2^{n/2}-1) (A
C - B^2)}\right) \ . \label{ei}
\end{equation}
We analyze the limit $n \rightarrow \infty$ for $s \ne 0.5$ and $s
= 0.5$ separately:

\bigskip

\emph{(i) $s \ne 0.5$}

\bigskip

In the limit of very high $n$ we can approximate the ground state
energy given in equation (\ref{ground}) by

\begin{equation}
E_-(s) \sim \frac{1}{2}\left(1-\sqrt{1-4 s (1-s)}\right) \ .
\label{approx}
\end{equation}
Therefore, the quantity

\begin{equation}
\alpha \sim \frac{1}{1 - \left(\frac{E_-(s)}{1-s}\right)}
\label{apal}
\end{equation}
diverges at $s=0.5$, which implies that this limit can not be
correct for that value of the parameter. 
 The closer we are to $s = 0.5$, the
bigger is $\alpha$. In this limit we find that

\begin{eqnarray}
A &\sim& \frac{\alpha^2 + 2^{n/2}}{\alpha^2 + 2^n}
\nonumber \\
&& \\
B &\sim& \frac{\alpha + 2^{n/2}}{\alpha^2 + 2^n}
\nonumber \\
&& \\
C &\sim& \frac{2^{n/2}}{\alpha^2 + 2^n} \ , \label{limit}
\end{eqnarray}
where all these quantities tend to zero as $n \rightarrow \infty$.
It is important to note that the convergence of the limit depends
on the value of $\alpha$ or, in other words, how close to $s=0.5$
we are. The closer we are to $s=0.5$, the slower is the
convergence, and therefore any quantity depending on these
parameters (such as the entropy) will converge slower to its
assimptotical value. For the eigenvalues of the reduced density
matrix we then find that when $n \rightarrow \infty$

\begin{equation}
\lambda_{\pm} \rightarrow \frac{1}{2} (1 \pm 1) \ , \label{eigenv}
\end{equation}
so $\lambda_+ \sim 1$ and $\lambda_- \sim 0$, and therefore the
assimptotical entropy is

\begin{equation}
E(s\ne 0.5, n\rightarrow \infty) = -\lambda_+ \log{ _2\lambda_+} -
\lambda_- \log{ _2\lambda_-} = 0 \ . \label{entropia}
\end{equation}
The convergence of this quantity is slower as we move towards $s =
0.5$.

\bigskip

\emph{(ii) $s = 0.5$}

\bigskip

We begin our analysis by evaluating the quantities at $s = 0.5$
and then taking the limit of big size of the system. We have that
$\alpha (s = 0.5) = \frac{2^n - 1}{2^{n/2}-1} \sim 2^{n/2}$. From
here it is easy to get the approximations

\begin{eqnarray}
A &\sim& \frac{1}{2} \nonumber \\
&& \nonumber \\
B &\sim& \frac{1}{2^{n/2}} \nonumber \\
&& \nonumber \\
C &\sim& \frac{1}{2^{n/2 + 1}} \ , \label{abc}
\end{eqnarray}
and therefore

\begin{equation}
\lambda_{\pm} \sim \frac{1}{2}\left(1 \pm \sqrt{1 - 4 \ 2^{n/2}
\left(\frac{1}{4} \frac{1}{2^{n/2}} - \frac{1}{2^n}\right)}\right)
= \frac{1}{2} \pm \frac{1}{2^{n/4}} \ , \label{eigensdos}
\end{equation}
so $\lambda_{\pm} \rightarrow \frac{1}{2}$ and $E(s=0.5, n
\rightarrow \infty) = 1$. According with (\ref{eigensdos}) we can
evaluate the finite size corrections to this behavior and find the
scaling of the entropy with the size of the system for very large
$n$. The final result for the entropy at the critical point reads

\begin{equation}
E(s=0.5, n >> ) \sim 1 - \frac{4}{\ln{2}} 2^{-n/2} \ .
\label{scaling}
\end{equation}
Note that the entropy remains bounded and 
tends to $1$ for $s = 0.5$ as an square root in the
exponential of the size of the system, which is the typical factor
in Grover's quantum algorithm.

\begin{figure}
\centering
\includegraphics[angle=-90, width=0.8\textwidth]{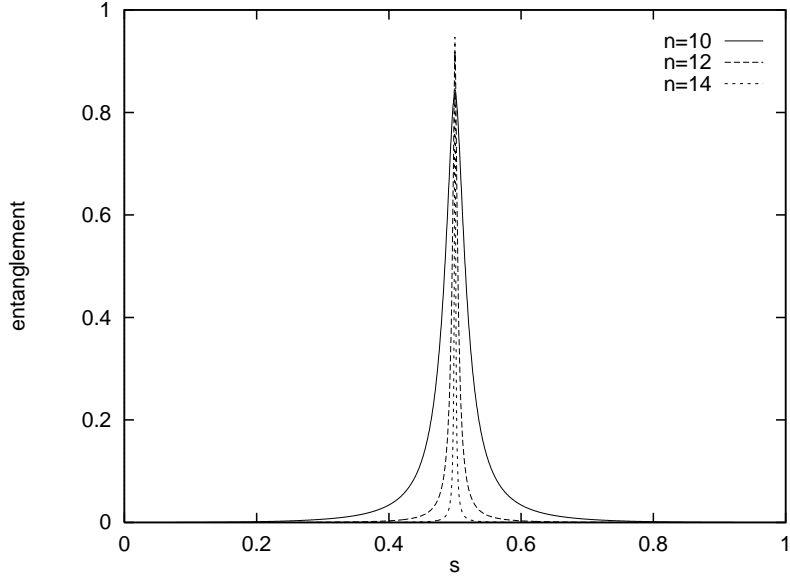}
\caption{Von Neumann entropy for the reduced system as a function
of $s$ for $10$, $12$ and $14$ qubits. As the size of the system
increases the entropy tends to zero at all points, except at $s =
0.5$ in which tends to $1$.} \label{pic}
\end{figure}
\begin{figure}
\centering
\includegraphics[angle=-90, width=0.8\textwidth]{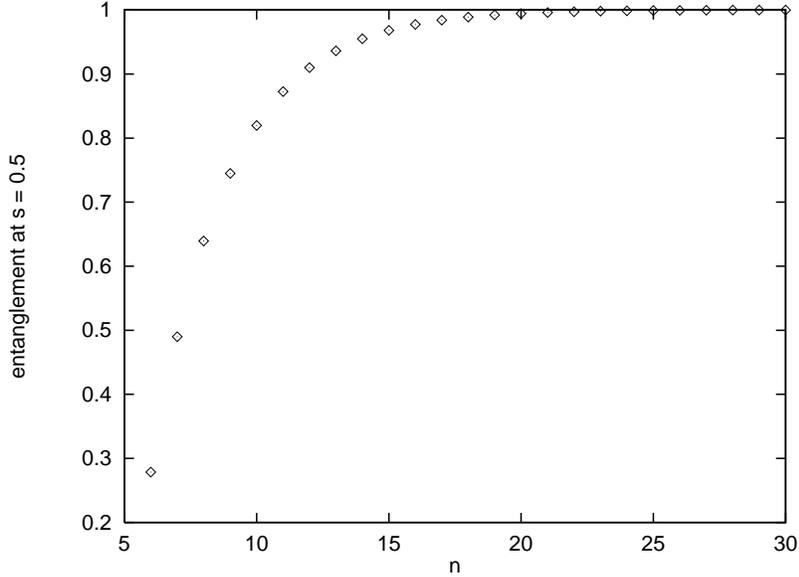}
\caption{Von Neumann entropy for the reduced system at $s = 0.5$
as a function of $n$. For infinite size of the system there is a
saturation at $1$.} \label{scale}
\end{figure}

We have represented the evolution of the entanglement entropy as a
function of $s$ for different sizes of the system in Fig. \ref{pic}
and have plotted in Fig. \ref{scale} the maximum value of the
entropy along the computation as a function of the size of the
system according to the expression given in equation
(\ref{scaling}). We can now compare the two plots with Fig. \ref{mix-300} and
Fig. \ref{ent} in the previous section.  The behavior for 
the entropy in
Grover's adiabatic algorithm is dramatically different to the one
observed in the NP-complete problem. 
Entanglement gets saturated in Grover's adiabatic
algorithm \emph{even at the critical point}, 
which is reminiscent of short ranged quantum correlations
in quantum spin chains \footnote{A somehow
similar situation is present in one-dimensional quantum spin
chains outside of the critical region, where the entanglement
entropy also reaches a saturation when increasing the size of the
system \cite{kike2}. 
Saturation does not appear in higher dimensional systems.}. 

Let us note that, in the limit of
infinite size, the quantum state in Grover's algorithm is separable
with respect to any bipartition of the system (and therefore not
entangled, as it is a pure state) for any $s$ except for $s =
0.5$. All the entanglement along the algorithm is concentrated at this
point, but this entanglement is still a bounded quantity and
actually equal to $1$. Consequently, a small amount of
entanglement appears essentially only at one point when the size
of the system is big, whereas the rest of the algorithm
needs to handle just separable states. We point out that these results
apply as well to the traditional discrete-time implementation of Grover's searching
algorithm, as the states between iterations are the same as in the adiabatic
version for discrete $s$ values. 

\section{Conclusions}

In this paper we have studied the scaling of the entanglement
entropy in several quantum algorithms. In particular, we have
analytically proven that Shor's factoring algorithm makes use of an
exponentially large amount of entanglement between the target
register and the source register after the modular exponentiation
operation, which in turn implies the impossibility of an efficient
classical simulation by means of the protocol of
ref. \cite{guifre}. Furthermore, we have provided numerical
evidence for a universal linear scaling of the entropy with the
size of the system together with a polynomially small gap in a
quantum algorithm by adiabatic evolution devised to solve the
NP-complete problem Exact Cover, therefore obtaining a
polynomial-time quantum algorithm which would involve exponential
resources if simulated classically, in analogy to Shor's algorithm.
Universality of this result follows from the fact that the quantum
adiabatic algorithm evolves close to a quantum phase transition and
the properties at the critical region do not depend on
particular details of the microscopic Hamiltonian (instance) such as
interactions among the spins or local magnetic fields. We have
also proven that the Von Neumann entropy remains a bounded
quantity in Grover's adiabatic algorithm regardless of the size of
the system even at the critical point. More concretely, the
maximum entropy approaches to one as an square root in the size of
the system, which is the typical Grover's scaling factor.

Our results show that studying the scaling of the entropy is a
useful way of analyzing entanglement production in quantum
computers. Results from other fields of physics
\cite{cft1,cft2,cft3} can be directly applied to bring further
insight into the analysis of quantum correlations. Different
entanglement scaling laws follow from different situations
according to the amount of correlations involved, as can be seen
in Table \ref{scalinglaws}. A quantum algorithm can be understood
as the simulation of a system evolving close to a 
quantum phase transition. The amount of entanglement
involved depends on the effective dimensionality of the system,
which in turn governs the possibilities of certain efficient
classical simulation protocols.
\begin{table}[h]
\begin{center}
\begin{turn}{90}
$\longleftarrow$ less entanglement
\end{turn}
\begin{tabular}{|c||c|}
    \hline
     & \\
   Problem & Entanglement scaling  \\
     & \\ \hline \hline
     & \\
   Adiabatic Exact Cover's quantum algorithm  & $E = O(n)$  \\
     & \\
   Shor's quantum factoring algorithm & $E = O(\log{_2 \ r}) \sim O(n)$  \\
     & \\
   Critical $d$-dimensional spin networks & $E = O(n^{\frac{d-1}{d}})$  \\
     & \\
   Critical one-dimensional spin chains & $E = O(\log{_2 \ n})$  \\
     & \\
   Non-critical one-dimensional spin chains & $E = O(1)$ \\
     & \\
   Adiabatic Grover's quantum algorithm & $E = O(1)$ \\
     & \\
    \hline
\end{tabular}
\end{center}
  \caption{entanglement scaling laws in different problems, 
in decreasing complexity order.}
  \label{scalinglaws}
\end{table}

These scaling laws provide  a new way of understanding some
 aspects from one-way quantum computation. It is known that
the so-called cluster state of the one-way quantum computer can be
generated by using Ising-like interactions on a planar
two-dimensional lattice \cite{hans1, hans2, hans3}.
This fact can be related to the linear (in the size of a box)
behavior of entropy for
spin systems in two dimensions.
One-dimensional models seem not to be able to efficiently create
the highly-entangled cluster state \cite{priv1}. Again,
this fact can be traced to the logarithmic scaling law 
of the entropy in spin chains which is insufficient
to handle the large amount of entanglement to carry out
{\sl e.g.} Shor's algorithm. Note also that $d\ge 3$ dimensional
systems bring unnecessarily large entanglement.

Quantum phase transitions stand as the more demanding systems
in terms of entanglement. They are very hard to simulate classically.
It is then reasonable to try to bring NP-complete problems to
a quantum phase transition setup, which quantum mechanics
handles naturally. 

\bigskip

\textbf{Acknowledgements:} We are grateful to A. Ac\'{\i}n,
J. Bergli, H. J. Briegel, A. Childs, M. A.
Mart\'{\i}n-Delgado, E. Farhi, L. Masanes, K. Pilch, E.
Rico and G. Vidal for discussions about the content of this
paper. We acknowledge financial support from the projects MCYT
FPA2001-3598, GC2001SGR-00065, IST-1999-11053, PB98-0685 and
BFM2000-1320-C02-01. Part of this work was done at the Benasque
Center for Science.

\end{document}